\documentclass[onecolumn,9pt]{article}
\pdfoutput=1
\usepackage{graphicx} 
\usepackage[english]{babel}
\usepackage{subcaption}
\usepackage{changepage}
\usepackage{authblk}

\title{Fluorinated surfaces: towards a universal matrix-free substrates for Laser Desorption Ionization}
\author[1]{Chiara Piotto}
\author[2]{Graziano Guella}
\author[1]{Paolo Bettotti}
\affil[1]{Nanoscience Laboratory, Department of Physics, University of Trento}
\affil[2]{Center for Agricultur, Food and Environment, Department of Physics, University of Trento}

\begin{document}
\maketitle

\begin{abstract}Matrix Assisted Laser Desorption Ionization (MALDI) is a soft ionization method that finds  widespread applications in high-throughput mass spectrometric analysis. One of the main limit of this technique is that it requires the use of matrices: these molecules enable the analytes  desorption and ionization (D/I) processes but  also generate  strong interfering signals  in the low mass spectral region, preventing a suitable detection  of low molecular weight compounds. The possibility to avoid their use will  ease both sample preparation and mass spectrum (MS) interpretation. In recent years nanostructured surfaces have been proposed as a viable method to achieve such goal but the results are limited to specific classes of chemical compounds and the approach lacks in generality. Here we demonstrate that the fluorination of surfaces is the only step needed to activate the target and to achieve matrix-free operation with the same detection limit obtained using nanostructured chips. The results of this work suggest that D/I is a purely interfacial effect, with negligible contribution from chemical nature of the underlying substrate, provided that the substrate is  conductive and able to efficiently absorb the UV laser beam.
\end{abstract}

\section{Introduction}
Laser Desorption Ionization (LDI) is a powerful technique exploiting electronic excitation to investigate medium- and even large-sized (bio)molecules as proteins. The possibility to integrate such ion-source with accurate ion-analyzers such as Time of Flight (TOF), enabled the development of the well known  MALDI-TOF mass spectrometric technique.\\
Due to its high sensitivity, buffers tolerance and very fast scan speed, MALDI-TOF is today a standard tool in mass spectrometry, in particular in proteins/peptides analysis. However, its application to a broader range of compounds belonging to different chemical classes is limited by four main issues: 
\begin{itemize}
\item MALDI requires the mixing of the analytes with a proper matrix, generally constituted by organic aromatic  acidic compounds, that provides protons able to assist and promote the desorption and ionization(D/I) processes of analytes.  The choice of the matrix  and the optimization of the molar ratio [matrix]/[analyte] are still found using empirical approaches.
\item The matrix molecules desorb together with the analytes and, being small-sized compounds,  they generate strong interfering signals at low m/z ratios, thus hindering the detection and structural characterization of low molecular weight analytes such as those within the range 150-500 Da. 
\item  The inhomogeneous crystallization of the matrix with the analyte produces a poor shot-to-shot and sample-to-sample signal reproducibility.
\item The nature of the ionization process  remains poorly understood and no single ionization mechanism is able to explain the complexity of the D/I mechanisms operating in MALDI.
\end{itemize}

A matrix-free LDI has a two-fold advantage: it requires a simpler sample preparation and it permits a  wider applicability with respect to MALDI, in particular in the low mass region\cite{art1paolo}. Several nanostructured substrates were proposed for matrix-free LDI measurements ranging from metals\cite{metals}, to carbon-based\cite{carbon}, to semiconductors and even insulators\cite{paa,Sunner}. The appearance of so many different active substrates supports the idea that  the D/I is driven by interfacial mechanisms, rather than determined by the physico-chemical nature of the underlying inorganic substrate.\\
In 1999 Wei and coworkers \cite{nat99} exploited silicon nanostructures to achieve matrix-free LDI: the technique was called  Desorption Ionization on porous Silicon (DIOS). 
Since DIOS has been proposed for the first time, hundreds of papers tried to understand which parameters optimize the D/I process: some researchers focused on substrates morphology, other on surfaces wettability (modified by proper functionalization) but contradictory results appeared in the literature on this topic (a detailed review of the published data is reported in ESI Tab. S1\dag). Curiously, some of the articles dealing with DIOS reported an increase of the efficiency of the porous silicon (PSi) samples functionalized with a specific fluorosilane ((3-pentafluorophenyl)-propyldimethylchlorosilane (PFPPDCS)), without proposing convincing reason for this effect.\cite{paper10,paper12,paper16,paper24}\\
Our results suggest a different interpretation of the experiments and "minimize" the importance of the nanostructures for the D/I processes. In fact we demonstrate that:
\begin{itemize}
\item [1.] despite the extensive literature,  PSi is not effective in the D/I process unless HF traces remain in its porous structure;
\item [2.] PSi substrates functionalized with PFPPDCS are efficient surfaces for D/I and all the tested molecules are detected at femtomolar concentration;
\item [3.] different type of PFPPDCS-functionalized \textbf{\underline{bulk}} materials (such as silicon (Si) and aluminum (Al)) are efficient matrix-free substrates and do not require nanostructuring. Furthermore the sensitivity achieved with bulk substrates is of the same order of the one reported for DIOS chips.
\end{itemize}

Points 1 and 2 are strongly related: while naked PSi is inert and does not promote D/I, as soon as it is functionalized with PFPPDCS, acceptable ionization and good sensitivity are readily achieved with no clear dependence on the PSi nanostructure. Furthermore the effect of HF traces remaining within the nanopores is easily checked and further supports the role of fluoride species in D/I, as underlined below. Both these resutls are hardly "justifiable" by a non-optimized PSi nanostructure or by oxidation-related effects.\\
Point 3 is a further demonstration that D/I is neither directly related with the chemistry of the underlying substrate nor it depends on the surface nanostructuring. Rather it is mainly driven by interfacial effects.\\
As analytes we tested different types of molecules: (Des-Arg9)-bradykinin and reserpine, since they have been analyzed in several DIOS studies; moreover we tested 3 relevant small biomolecules (abscisic acid (ABA), luteolin and quercetin). ABA is an important plant hormone, involved in many plant developmental processes. Worth of note it is a regulator of various environmental stress responses \cite{guella1}.  Even quercetin and luteolin are important bioactive compounds with antioxidative and anti-inflammatory properties;  they are the flavonoids more widely distributed in nature, mainly in fruits, vegetables, and teas \cite{guella2}. The detection and quantification of these small-sized metabolites in real biological samples (plant extracts) is still an analytical challenge. This is the main reason that prompted us to choose these molecules to test our methodology.

\section{Experimentals}
\subsection{Etching of Porous Silicon samples} 
PSi samples were fabricated with the usual electrochemical etching\cite{psietching}. We prepared hundreds of morphologically different PSi (see ESI Tab. S2-S4\dag). After the etching, all samples were thoroughly rinsed in ethanol and dried under nitrogen flow, with the exception of those containing HF traces; on the latter the analytes were deposited immediately after the etching, without any rinsing. 

\subsection{Samples oxidation and functionalization} 
Some PSi samples have been used as prepared (fresh PSi); other have been either oxidized or functionalized.  
PSi chips have been oxidized in different ways: a thin oxide layer spontaneously growth if the samples were let in air for 12 h; thermal oxide with different thickness have been growth  by letting the samples at ambient atmosphere either at 300$^{\circ}$C for 7 h or at 150$^{\circ}$C for 3 h. Finally wet oxidation has been achieved by dipping the samples in deionized water for 2h.\\ 
To silanize the PSi substrates with PFPPDCS, they  were dipped in water for 12 h followed by their drying through a thermal treatment (2.5 h, 90 $^{\circ}$C). 10 $\mu l$ of PFPPDCS were deposited on the samples and left on an hot plate at 90 $^{\circ}C$ for 2 h. All samples were rinsed with ethanol to remove the  excess of the silane-reagent.\\
Silicon, aluminum  and indium tin oxide bulk samples (F-bulk-Si, F-bulk-Al, F-bulk-ITO) were silanized with PFPPDCS: after 2 min air plasma cleaning, 10 $\mu l$ of PFPPDCS were deposited on each sample and left on an hot plate (90 $^{\circ}$C) for 2 h. All samples were rinsed with ethanol to remove the excess of the silane-reagent.

\subsection{Tested Analytes and Analytes preparation}
Various analytes were tested: (Des-Arg9)-bradykinin (Tocris Bioscience), reserpine, luteolin, quercitin and abscisic acid (all from Sigma-Aldrich).\\
To perform the MALDI experiments the analytes were mixed with 2,5-dihydroxybenzoic acid (DHB, Sigma-Aldrich; [analyte]:[matrix] = 1:5000) and methanol (Sigma-Aldrich) to obtain the following concentrations : (Des-Arg9)-bradykinin 1 ng/$\mu l$, reserpine 0.6 ng/$\mu l$, luteolin and quercitin 500 ng/$\mu l$, ABA 190 ng/$\mu l$.\\
To perform matrix-free LDI, the targeted analytes were dissolved in various solvents (methanol; H$_{2}$O:methanol 1:1; 0.1\% aqueous TFA:methanol 1:1; aqueous NaCl 1.35 mM:methanol 1:1).  LDI measurements were performed at different concentrations (1000 ng/$\mu l$, 500 ng/$\mu l$, 250 ng/$\mu l$, 100 ng/$\mu l$, 10 ng/$\mu l$, 1 ng/$\mu l$ 0.6 ng/$\mu l$ and 0.1 ng/$\mu l$).\\

\subsection{MALDI and matrix-free LDI analysis}
We used a MALDI-TOF instrumentation (Ultraflex III TOF/TOF, Bruker; N$_{2}$ laser with 3 ns pulse at 337 nm; maximum energy/pulse: 93 $\mu J$; repetition rate of 20 Hz). Both the energy per pulse (set at about 65 $\mu J$) and the number of laser shots were adjusted during the measurements to find optimal conditions. The extraction field for cations was -25 kV, whereas for anions was +20 kV. The extraction time was kept fixed at 100 ns. All the analyses were performed in the reflector mode and a suitable external calibration was performed before each series of measurements.

\section{Discussion}
\begin{figure}
\centering
\includegraphics[width=0.8\columnwidth]{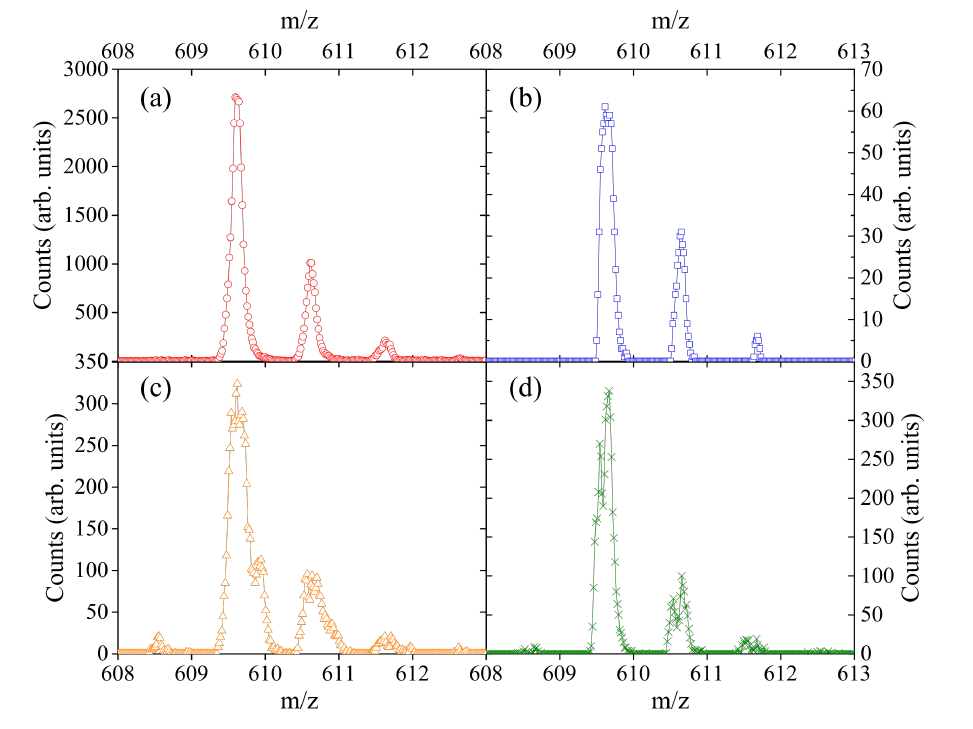}
\caption{LDI-MS of reserpine (isotopic cluster of [M+H]$^{+}$ peak at m/z 609.6 Da) obtained on different substrates: a) PSi+HF; b) F-PSi; c) F-bulk-Si; d) F-bulk-Al.} \label{fig1}
\end{figure}

We tested hundreds of PSi samples to check their efficiency as matrix-free LDI substrates (a detailed list is found in ESI Tab. S2-S4\dag). Despite the large number of different porous morphologies and the different extent of surface oxidation of these substrates, none of the tested analytes was detected   neither in positive nor in negative ion-mode. This outcome suggests that the requirement of a nanostructured surface as LDI substrate is not sufficient to achieve an efficient D/I process of our molecules. As a remarkable finding, when PSi samples were not thoroughly rinsed with aqueous ethanol to remove HF traces, all analytes were detected with the same femtomole sensitivity reported in state-of-the-art literature (see ESI S3\dag for the details of sensitivity estimation). Among the various tested compounds, reserpine has been chosen as a model molecule and its MS acquired using PSi+HF chip is shown in Fig. \ref{fig1}(a).

Considering the fact that the analytes are detected as the corresponding [M+H]$^{+}$ protonated species (except for ABA, that was detected as [M+Na]$^{+}$), we investigated if the presence of H$^{+}$  ions within the pores was sufficient to induce analytes D/I. After PSi samples preparation and before analyte deposition, we added a small amount of    weak (aq. CH$_{3}$COOH) and/or strong (aq. HCl) acids, but none of the target analytes were succesfully detected on acidified substrates. Therefore our measurements indicate that HF does not play a major role as a proton transfer chemical species but rather it acts as a source of F$^{-}$ anion, with the latter apparently able to assist the analytes D/I. The fact that fluoride species have an active role in D/I is confirmed by the experiments performed using PFPPDCS functionalized surfaces. 

According to  literature\cite{paper10,paper12,paper16,paper24}, PFPPDCS seems able to increase the efficiency of the PSi samples; in fact when we used florinated PSi chips (F-PSi) all the tested molecules were detected (an example of the Reserpine MS is reported in Fig. \ref{fig1}(b)).\\
These results led us to suppose that the fundamental role for the D/I processes is played by the  fluorinated silane rather than the presence of nanostructures within the substrate. Thus, our results are in sharp contrast with literature findings where the nanostructures are thought to be crucial for D/I processes\cite{laserablation, thermalc2}.

To test our hypothesis we functionalized bulk-Si substrates: the bulk-Si chips functionalized with PFPPDCS (F-bulk-Si) were able to assist the LDI process as much as the F-PSi substrates and with comparable femtomoles sensitivity (see Fig. \ref{fig1}(c)).  As for the F-PSi cases all the analytes (except ABA) were detected as the corresponding [M+H]$^{+}$ protonated species. This finding clearly demonstrates that the nanostructuring is not required for D/I processes once the silicon substrates are functionalized with PFPPDCS. The control experiment, using the naked-bulk-Si substrates, confirmed that none of the targeted compounds could be detected.

\begin{figure}
\centering
\includegraphics[width=0.8\columnwidth]{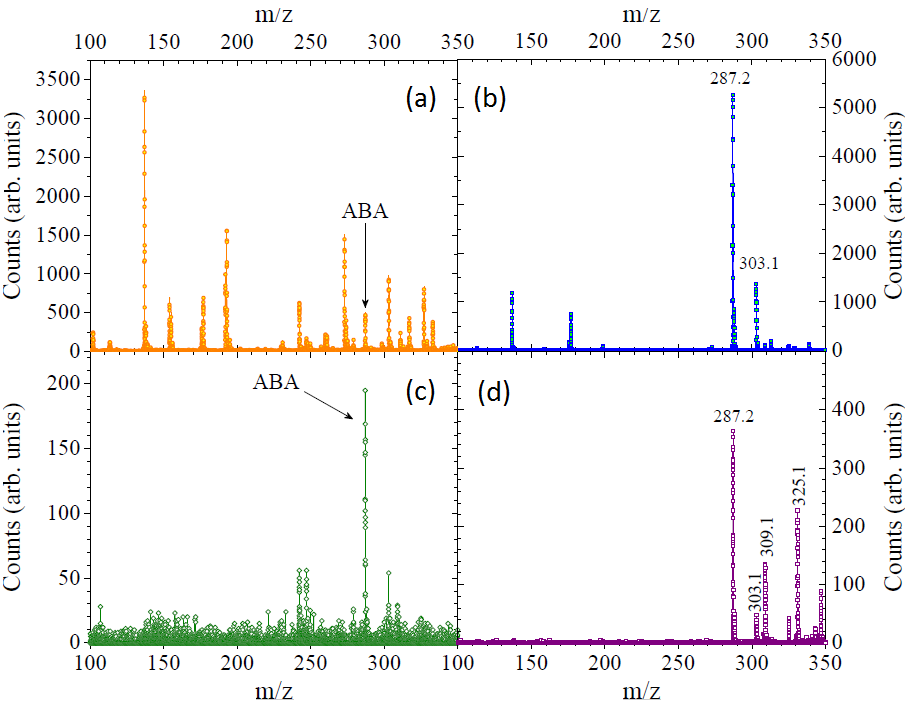}
\caption{a) ABA MALDI MS; b) 1:1 mixture of quercetin and luteolin  MALDI MS; c)ABA F-bulk-Si MS; d)1:1 mixture of quercetin and luteolin F-bulk-Si MS.}\label{fig3}
\end{figure}

To investigate if the electronic structure of the underlying substrate plays a fundamental role for molecules D/I,  we silanized bulk-aluminum  substrates. Also in this case all the analytes have been detected  (see Fig. \ref{fig1}(d)), thus the chemical nature of the substrate is not a key element to achieve matrix-free operations and is irrelevant for the D/I mechanisms.\\
On the other hand, the lack of signal from ITO samples silanized with PFPPDCS suggests that LDI processes are  thermally activated (see ESI Sec. S5\dag for a more in depth discussion). In fact ITO is a conductive material but does not absorbs the UV laser. This fact confirms the well known rule that matrix-free LDI substrate should be both conductive material as well as an efficient UV absorbers.\\
Among the various tested molecules, the (MA)LDI-TOF measurements on 3 relevant small biomolecules (ABA, luteolin and quercetin) are also reported.\\
In the typical MALDI spectra (as reported in Fig. \ref{fig3}(a,b))  the low m/z region is crowded of peaks arising from the matrix (in our case DHB)   that complicate the   analysis; instead when  using F-bulk-Si  substrates, the only relevant peaks are those generated from the analytes (see Fig. \ref{fig3}(c,d)).

Since PSi, bulk-Si and bulk-Al chips become all efficient substrates for matrix free MALDI once fucntionalized with PFPPDCS, an obvious question arises:  what is the role of the fluoride species in D/I?\\
First of all we excluded that this silane could play any role  in UV laser absorption: its molar absorption coefficient at 337 nm was estimated (see ESI Fig. S1\dag) to be about 5 L mol$^{-1}$ cm$^{-1}$, thus 3 orders of magnitude lower than that of typical MALDI matrices (the 2,5-Dihydroxybenzoic acid (DHB) molar absorption coefficient at 337 nm is $5\cdot 10^{3} L mol^{-1} cm^{-1}$)\cite{tesi35}.\\
Furthermore, by irradiating a F-bulk-Si chip (without depositing any analyte) no signal were obtained neither in positive nor negative mode, thus PFPPDCS is not ionized directly by laser irradiation.\\
By performing contact angle (CA) measurement on a macroscopic area scanned by the laser, we have verified that the silane was still attached to the chip surface after LDI analysis, as the CA remain constant to 82$ ^{\circ}\pm$3$ ^{\circ}$ (the CA measured on naked-bulk-Si surface was  33$ ^{\circ}\pm$3$ ^{\circ}$).\\
Additionally the root mean square (RMS) roughness obtained from AFM measurements on the area shot by the laser (0.7$\pm$0.2 nm) was compatible with that measured on the not-irradiated region (0.9$\pm$0.2 nm). Since a naked-bulk-Si substrates shows an RMS roughness of 0.3$\pm$0.1 nm, we concluded that:
\begin{enumerate}
\item the silane molecules are not ionized upon laser irradiation;
\item they are not detached as neutral species either.
\end{enumerate}
Therefore we excluded any direct interaction between the silane and the laser beam. We cannot exclude photochemical reaction paths when both silane and target molecules were both present on the chip but this hypothesis is hardly justified considering the different chemistry involved when diverse molecules were investigated.

\section{Conclusions}
In this article we propose the use of fluorinated surfaces as matrix-free substrates for LDI analysis. We detect several molecules with good sensitivity, on different target materials. Our findings demonstrate that, despite the extensive bibliography, as etched, naked-PSi chips are not efficient substrates for mass spectrometric analyses. Only if HF traces are present within the nanopores the analytes undergo efficient D/I processes. We added different acids on the naked-PSi chips but only the presence of HF traces allowed an efficient detection of all the tested analytes. This outcome clearly speaks for an active and essential role played by the F$^{-}$ anion in the D/I processes.\\
The fact that only HF traces allowed for the analytes detection  suggests that the F$^{-}$ anion has a role in the analytes D/I. Our finding cannot be ascribed to a poorly nanostructured PSi because as soon as we functionalized it with the silane we obtained high and reproducible sensitivity.\\
Moreover our experiments indicate that the role of the nanostructures has been greatly overestimated so far: not only the F-PSi chips, but also silicon and aluminum PFPPDCS-bulk substrates can be used for mass spectrometric applications and the sensitivity of the bulk materials is comparable to that of the nanostructured counterparts.

\section*{Acknowledgments}
This work was partially supported by the Italian Ministry of University and Research through the "Futuro in Ricerca" project RBFR12OO1G-NEMATIC. The authors thank Mr. A. Sterni for the technical support and Prof. M. Scarpa for fruitful suggestions.

\newpage
{\centering
\section*{Supplementary Information}
}

\section{Literature review of the key parameters for DIOS processes}
\begin{table}[!h]
  \caption{Literature survey and conflicting claims of some of the most cited DIOS-related articles.}
    \label{tb1}
  \begin{tabular}{|c|p{2cm}|p{2cm}|p{2cm}|p{2cm}|p{2cm}|}
    \hline
    Ref. & Pore Depth & Pore Diameter & Roughness & Interpore distance &  Porosity \\
    \hline
    \cite{nat99bis}&$\mu m$ thick &micro- and meso-pores& &&\\
    \hline
    \cite{tesi46}&best result: 5 $\mu m$, signal up to 8 $\mu m$& not crucial& best result: RMS=200 nm&&\\
    \hline
    \cite{tesi47}&up to 200 nm&70-120nm&&100 nm& 30-40\%\\
    \hline
    \cite{tesi48}&hundreds of nm&$\mu m$ range&&&\\
    \hline
    \cite{tesi49}&signal from 10nm, best results: >250 nm&&Signal also with non-porous substrates: RMS is the most important parameter&&\\
    \hline
    \cite{tesi50}&250-300 nm&30nm; no signal if diam < 3 nm&&&\\
    \hline
    \cite{tesi51}&< 5 $\mu m$&not crucial &&&\\
    \hline
    \cite{tesi52} & 200-460 nm, deeper pores reduce laser threshold & 100-600 nm && 16-960 nm & 4-92\% lower threshold at higher porosity\\
    \hline
    \cite{tesi53}& 200-500 nm & 2-50 nm&&&\\
    \hline
  \end{tabular}
\end{table}

\section{Parameters changed during the preparation of PSi samples}
Tables S2-S4 list the parameters varied during the preparation of the PSi samples (Si wafer resistivity and doping, HF concentration and solvent, current density and etching time). N-type PSi samples have been etched also with different illumination conditions: front-, back-illuminated or without light.\\
Once the PSi samples have been fabricated, various storage conditions have been adopted: some samples have been prepared few minutes before the experiment, others have been stored for different times (between 1 to 7 days) in ethanol, toluene, water or air. Also thermal oxidation has been performed, leaving the samples for some hours (from 3 to 7 h) on a hot plate at 70$^{\circ}$C.

\begin{table}[!h]
\caption{PSi samples prepared on n-type Si 0.01 $\Omega cm$.}
\label{table:tb2}
\begin{adjustwidth}{-1.5cm}{-1cm}
\begin{tabular}{|c|c|c|c|c|c|c|c|c|c|c|c|c|c|p{2cm}|}
    \hline
    &\multicolumn{6}{c|}{HF \% in ethanol}&\multicolumn{4}{c|}{HF \% in water}&\multicolumn{3}{c|}{HF \% in DMSO}&4\% HF in water:DMSO =1:10v/v\\
    \hline
    mA/cm$^{2}$&2.5&5&16&20&25&37.5&2.5&5&6.25&10&11.25&16&20&\\
    \hline
    2&&&&&&&&&&&&&&\\
	\hline
    5&&&&&X&&X&&&&&&&\\
    \hline
    10&&&X&&&X&&&&&&X&X&\\
    \hline
    15&&&X&&&X&&X&&&&&&\\
    \hline
    20&&&X&&X&X&&&X&&&X&X&\\
    \hline
    30&&&X&&&&&&X&&&&X&\\
    \hline
    35&&&X&&&&&&&&&&&\\
    \hline
    40&&&X&&&&&&X&&&&&\\
    \hline
    50&&&&&&&&&&&&&&\\
    \hline
    60&&&&&&&&&&&&&&\\
    \hline
  \end{tabular}
\end{adjustwidth}
\end{table}

\begin{table}[!h]
\caption{PSi samples prepared on p-type Si 0.002 $\Omega cm$.}
\label{table:tb3}
\begin{adjustwidth}{-1.5cm}{-1cm}
  \begin{tabular}{|c|c|c|c|c|c|c|c|c|c|c|c|c|c|p{2cm}|}
    \hline
    &\multicolumn{6}{c|}{HF \% in ethanol}&\multicolumn{4}{c|}{HF \% in water}&\multicolumn{3}{c|}{HF \% in DMSO}&4\% HF in water:DMSO = 1:10v/v\\
    \hline
    mA/cm$^{2}$&2.5&5&16&20&25&37.5&2.5&5&6.25&10&11.25&16&20&\\
    \hline
    2&&&&&&&&&&&&&&\\
	\hline
    5&&&&&X&X&&&&&&&&\\
    \hline
    10&&X&X&X&&X&&&&&&&X&\\
    \hline
    15&X&&X&&&X&&&&&&&&\\
    \hline
    20&X&&X&X&&X&&&&&&&X&\\
    \hline
    30&&&&X&&&&&&&&&X&\\
    \hline
    35&&&X&&&&&&&&&&&\\
    \hline
    40&&&&&&X&&&X&&&&&\\
    \hline
    50&&&&&&&&&&X&&&&\\
    \hline
    60&&&&&&X&&&&&&&&\\
    \hline
  \end{tabular}
\end{adjustwidth}
\end{table}

\clearpage

\begin{table}[!h]
\caption{PSi samples prepared on p-type  Si 3-6 $\Omega cm$.}
\label{table:tb4}
\begin{adjustwidth}{-1.5cm}{-1cm}
  \begin{tabular}{|c|c|c|c|c|c|c|c|c|c|c|c|c|c|p{2cm}|}
    \hline
    &\multicolumn{6}{c|}{HF \% in ethanol}&\multicolumn{4}{c|}{HF \% in water}&\multicolumn{3}{c|}{HF \% in DMSO}&4\% HF in water:DMSO = 1:10v/v\\
    \hline
    mA/cm$^{2}$&2.5&5&16&20&25&37.5&2.5&5&6.25&10&11.25&16&20&\\
    \hline
    2&&&&&&&&&&&&&&X\\
	\hline
    5&&&&&&&&&&&&&&X\\
    \hline
    10&&&X&&&&&&&&X&&X&X\\
    \hline
    15&&&&&&&&&&X&&&&\\
    \hline
    20&&&X&&&&&&&&X&&X&\\
    \hline
    30&&&&&&&&&&&X&&X&\\
    \hline
    35&&&&&&&&&&&&&&\\
    \hline
    40&&&&&&&&&&&X&&&\\
    \hline
    50&&&&&&&&&&&&&&\\
    \hline
    60&&&&&&&&&&&&&&\\
    \hline
  \end{tabular}
\end{adjustwidth}
\end{table}

\section{Sensitivity Estimation}
A rough estimation of the detection limits obtained on the different substrates has been performed considering the area of the released drop (1 mm in diameter) and the amount of analyte deposited (0.1 ng/$\mu l$). By assuming homogeneous analyte dispersion, all the mentioned substrates can be used to detect analytes down to femtomoles level.

\section{Thermal confinement}
Both porous and bulk silicon efficiently absorb UV light; their absorption coefficients in the UV region are 10$^6$   and 10$^5$ cm$^{-1}$ respectively and between 300 and 900 K are independent on the temperature \cite{laserablation}.\\
The main difference between PSi and bulk silicon has to do with the energy dissipation by heat conduction, that is very efficient in bulk substrates (with thermal conductivity of 150 W/(m K)).
In PSi the efficient energy deposition (due to high absorption coefficient) along with a limited dissipation through heat conduction (thermal conductivity less than 5 W/(m K) \cite{Timosheno}) imply that the temperature rises more rapidly and to higher values than at the surface of bulk substrates.\\
The time constant for heat conduction ($\tau_{th}$) \cite{papere} can be calculated for both PSi and bulk substrates: $\tau_{th}(PSi)$ is of the order of some nanoseconds, instead $\tau_{th}(bulk)$ assumes values of the order of hundreds of picoseconds.\\
These values should be compared with the 3 $ns$ laser pulse duration ($\Delta t$). If $\tau_{th} \sim \Delta t$ (as in the case of PSi chips) the heat conduction is not important as an energy loss process and a relatively homogeneous temperature distribution within the excited volume is expected. This very rapid heating of the sample by the laser radiation resulting from a comparably slow heat conduction is known as "thermal confinement regime" and it generates a thermoelastic pressure pulse in the absorbing sample volume which travels out of the excited volume at the speed of sound, carrying away part of the deposited energy.\\
In the case of bulk silicon, $\tau_{th} < \Delta t$: the thermal conductivity is so high that the heat dissipation through thermal pathways is very efficient.\\
Since in PSi thermal confinement is achieved, several articles suggest that this is the reason why PSi can be exploited as a substrate in mass spectrometry \cite{laserablation, thermalc2}. Our results indicate that the thermal effect might have an important role in LDI (considering that the only inactive substrate is ITO, which is the only one transparent to the laser) but it is not necessary to achieve the very high temperatures obtained in PSi, rather a moderate temperature increase,obtained using opaque targets, is sufficient to generate LDI processes.

\section{PFPPDCS absorption spectrum}
\begin{figure}[!h]
\centering
\includegraphics[width=0.8\columnwidth]{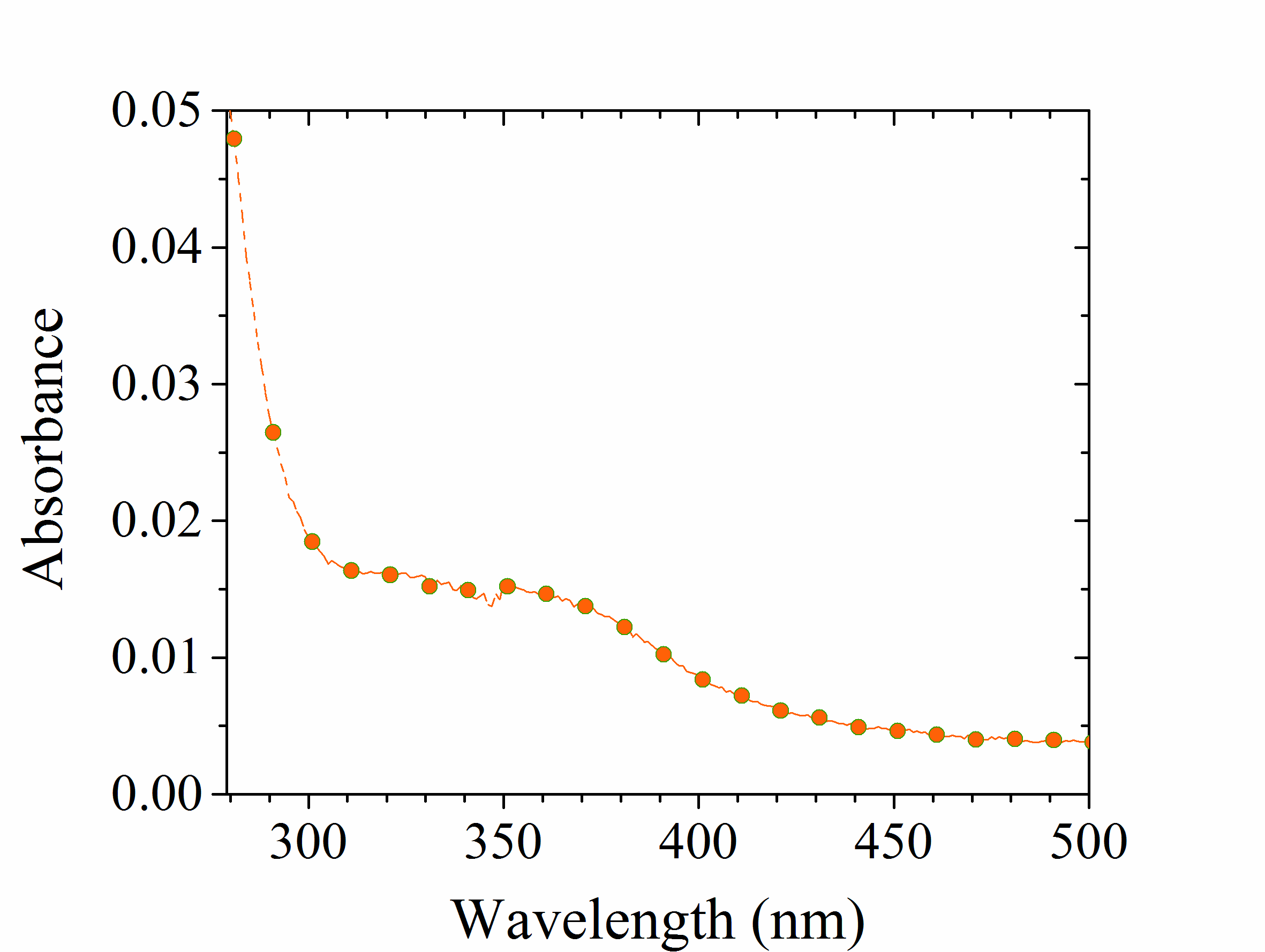}\\
\caption{A 30 mM PFPPDCS solution in methanol has been used to acquire the silane absorption spectrum using the Agilent Cary 5000 Spectrophotometer. The spectrum has been collected in transmittance mode, performing a baseline correction with a quartz cuvette full of methanol.}
\label{figS1}%
\end{figure}

\end{document}